\documentclass[twocolumn,showpacs]{revtex4}
\usepackage{array}
\usepackage{graphicx}
\usepackage{amsmath,amssymb}
\usepackage{dcolumn}
\usepackage{bm}
\begin{document}
\title{Pump process of the rotatory molecular motor and its energy
efficiency}
\author{Hiroshi Miki, Masatoshi Sato, and Mahito Kohmoto \\
\it The Institute for Solid State Physics, The University of Tokyo,\\
\it 5-1-5 Kashiwanoha, Kashiwa, Chiba 277-8581, JAPAN}
\date{\today}
\begin{abstract}
The pump process of the ratchet model inspired by the $F_o$ rotatory motor
 of ATP synthase is investigated. In this model there are two kinds of
 characteristic time. One is dynamical, the relaxation time of the
 system. Others are chemical, the chemical reaction rates at which a
 proton binds to or dissociates from the motor protein. The inequalities
 between them affect the
 behavior of the physical quantities, such as the rotation velocity and
 the proton pumping rates across the membrane. The energy transduction 
 efficiency is calculated and the condition under
 which the efficiency can become higher is discussed.
 The proton pumping rate and the efficiency have a peak
 where a certain set of
 inequalities between the chemical reaction rates and the reciprocal of
 the relaxation time holds. The efficiency
 also has a peak for a certain value of the load. The best efficiency 
 condition for the pump process is consistent with that for the
 motor process. 

\end{abstract}


\maketitle

\section{introduction}

Biomolecular motors are kinds of protein molecules which generate a
unidirectional motion in the situation that many surrounding molecules, 
{\it e.g.} water molecules, collide with the protein molecule many times 
due to thermal fluctuation\cite{How,VM}. They play many essential roles 
in life; muscle contraction, transport in cells, {\it etc}. It is 
interesting and  important question how motor proteins generate a
 unidirectional motion. It is natural to think that they  utilize this 
random fluctuation by the collision,
 although there has been no unambiguous evidence
 reported so far that this is essential.

Ratchet systems, which take out a net finite current from random and 
undirected noise, have been intensively investigated for modeling 
molecular motors\cite{Rats}. As known by the famous Feynman's
lecture\cite{Feynman} and many investigations
thereafter\cite{JAP,Reimann,Appl,AH}, a net finite directed 
motion is the consequence of the violation of the symmetry of the system,
 {\it e.g.} the spatial form of the potential. In modeling molecular
 motors, the system also needs to violate the detailed balance between
 several different states, {\it i.e.} conformations. The detailed 
balance is violated by the undirected chemical energy supply, {\it e.g.} 
the ATP hydrolysis. Ratchet systems are applied to various
physical systems other than modeling molecular
motors\cite{Reimann,Appl,AH}.

The $F_o$ part of the ATP synthase is known as a rotatory molecular energy
transducer\cite{VA,Sam,TAYC,RG,MLH}. It is embedded in the membrane, and
transduces the energy of the transmembrane proton concentration gradient, 
when a proton passes through the membrane via the $F_o$ part,  
into that of the rotation of its internal subunits in the motor process. 
The ratchet model of it was presented qualitatively by Junge, 
Lill and Engelbrecht\cite{JLE}. Elston, Wang, and Oster improved this 
model by taking account of the electrostatic interaction between the residues
\cite{EWO}. This ratchet model has an advantage that the coupling between
 the rotation and the proton flow is easy to see, therefore the energy transduction 
efficiency is easily estimated. In this model, there are two important kinds of
characteristic time, dynamical and chemical. The former is the relaxation time of the
convection-diffusion system. The latter are the chemical reaction
rates at which a proton binds to or dissociate from the motor protein
when it passes  through the membrane. We investigated the efficiency in detail 
and pointed out that in the motor process, the inequalities between them
affect  the behavior of it \cite{MSK}. For the motor to work most
efficiently, certain inequalities between them need to hold. 

The $F_o$ part also works as a pump\cite{spt}; it gains energy from the 
 (inverse) rotation and 
pumps out protons against the transmembrane proton gradient. 
Which process this protein adopts, motor or pump, under a given condition 
and what controls this determination are important questions, but satisfying
answers to these have not been obtained so far.  

In this paper we investigate the physical quantities in the pump process, 
such as the rotation velocity, the proton pumping rate, and the energy 
transduction efficiency. The condition under which the protein works as a
pump with  higher efficiency is discussed. We will compare the most
efficient  condition for the pump process with that for the motor process. 

\section{model}
\begin{figure}
\begin{center}
\includegraphics[width=7.5cm]{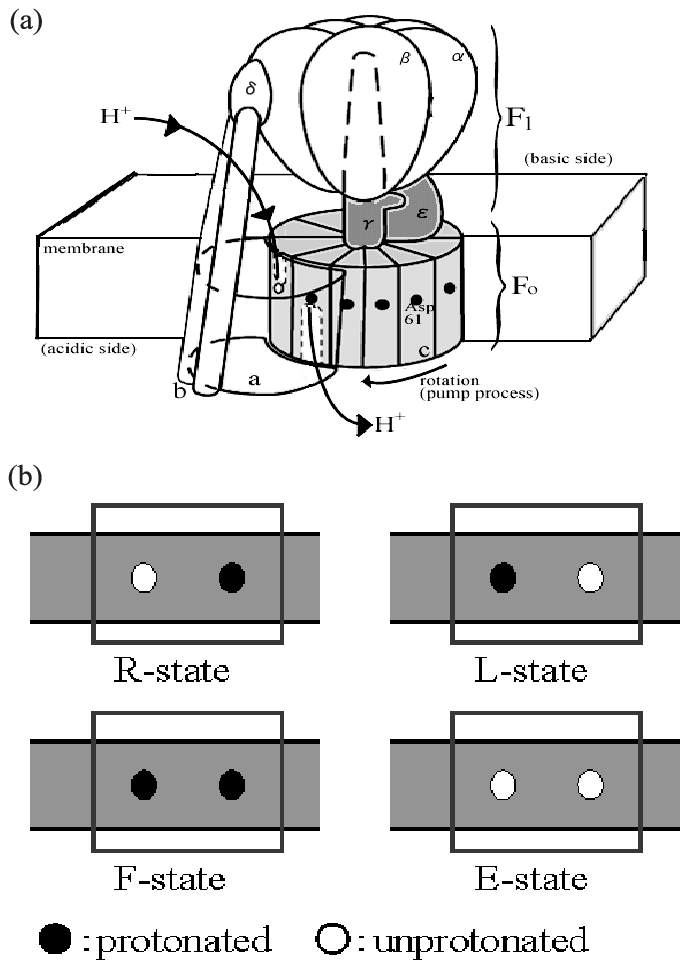}
\caption{The upper figure (a) is the ATP synthase. When it
 operates, the shaded parts ($\gamma,\epsilon$, and the {\bf c}-ring)
 rotate. In the pump process,  the external load  force from $F_1$
 rotates the {\bf c}-ring in the direction of the arrow and pumps
 out a proton from the upper to the lower side. This direction of
 rotation is opposite to that in the motor process. In the lower
 figures, (b), the states of the {\bf c}-ring are shown. The rectangle
 frame is the proton channel viewed from the {\bf a}-subunit and the
 shaded region is the {\bf c}-ring. 
The states depend on the protonations of the two proton binding sites 
 in the channel. The figure (a) corresponds to the
 R-state. \label{FO}}
\end{center}
\end{figure}

The ATP synthase is schematically shown in FIG.\ref{FO}(a). 
The upper side of the figure is the basic (inner) side of the cell
and the lower side is the acidic (outer) side. The proton
concentration  of the acidic side is 
kept higher than that of the basic side by respiratory chains. 
The $F_o$ part is composed of the subunits denoted by the Roman letters,
{\bf a}, {\bf b}, and {\bf c}, and the $F_1$ part by the Greek letters,
$\alpha$, $\beta$, $\gamma$, and $\delta$.
The rotating ring (the {\bf c}-ring)is composed  of the assembly of
the {\bf c}-subunits, each of 
which has a proton binding site, carboxyrate Asp61, almost in the middle 
of itself. The {\bf a}-subunit is fixed in the membrane. The proton
channel is the interface between the {\bf a}-subunit and the {\bf
c}-ring. There are two {\bf c}-subunits and two proton paths in 
the channel. One of the paths connects the left binding site of the
 {\bf c}-subunits to the basic side and 
the other connects the right binding site to the acidic side. The other
{\bf c}-subunits are in the membrane.  The proton 
binding site can be both protonated and unprotonated only if it is in the 
channel. Otherwise it can be only protonated due to the hydrophobicity 
of the membrane. The conformational  change of the {\bf c}-subunits between 
the two states, protonated and unprotonated, is small\cite{RG} and we
neglect the effect on the place of the binding sites. The states of the 
{\bf c}-ring are determined by protonations of the two proton binding sites in 
the channel; the empty(E) state with the both binding sites unprotonated, 
the right(R) state with the right binding site protonated and the left 
binding site unprotonated, the left(L) state with the left binding site 
protonated and the right binding site unprotonated, and the full(F) 
state with the both binding sites protonated. As a motor the {\bf c}-ring 
rotates rightward and as a pump leftward in FIG.\ref{FO}. The $\gamma$-
subunit connects the $F_o$ part to the $F_1$ part.  

We investigate this system by using the 'simply biased diffusion model' 
\cite{JLE,EWO, MSK}. It works as follows: The {\bf c}-ring moves randomly 
by collision with surrounding molecules and is forced to rotate by the 
external load from the $F_1$ part. But the hydrophobic interaction prevents 
the unprotonated binding site from moving into the membrane, {\it i.e.} the 
E and R states cannot move leftward out of the channel and the E and L 
states not rightward. If the chemical reaction rates, at which a proton 
binds to or dissociates from the binding sites in the channel, are
tuned appropriately, it becomes 
possible to take out a net motion against the load or to pump protons 
against the concentration gradient.   

We consider the pump process of this system. A proton passes through 
the channel from the basic side to the acidic side as follows: 
First a proton in the basic side flows into the channel and binds to the 
left binding site. Next the proton goes through  the membrane with the 
leftward rotation of the {\bf c}-ring. Finally the proton comes back to
the channel, 
dissociates from the right binding site, and flows out to the acidic side. 
In this way, a proton is pumped out by the rotation of the {\bf c}-ring 
driven by the external load, which is gained by the energy of the ATP 
hydrolysis in the external part. We assume that in the pump process, the 
external load exerted is larger than that in the motor process, so that
the {\bf c}-ring can rotate leftward. In this sense, the external 'load'
should be exactly called 'driving force' of the rotation in the pump
process. However, we will use the term 'load' in the pump process  since
we deal with the same model as in the motor process. 

\begin{figure}
\begin{center}
\includegraphics[width=7.5cm]{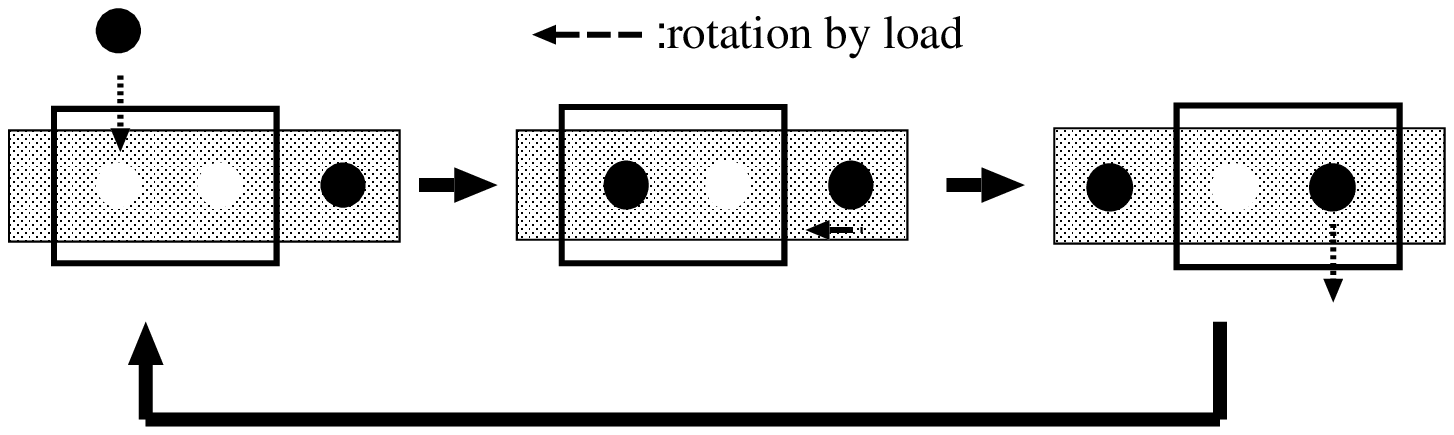}
\caption{The good process of the coupling between the proton pumping and
 the motion of the {\bf c}-ring. In this process, one step of the
 rotation tightly couples to one proton pumping. The energy is
 transduced efficiently. \label{good}}
\includegraphics[width=7.5cm]{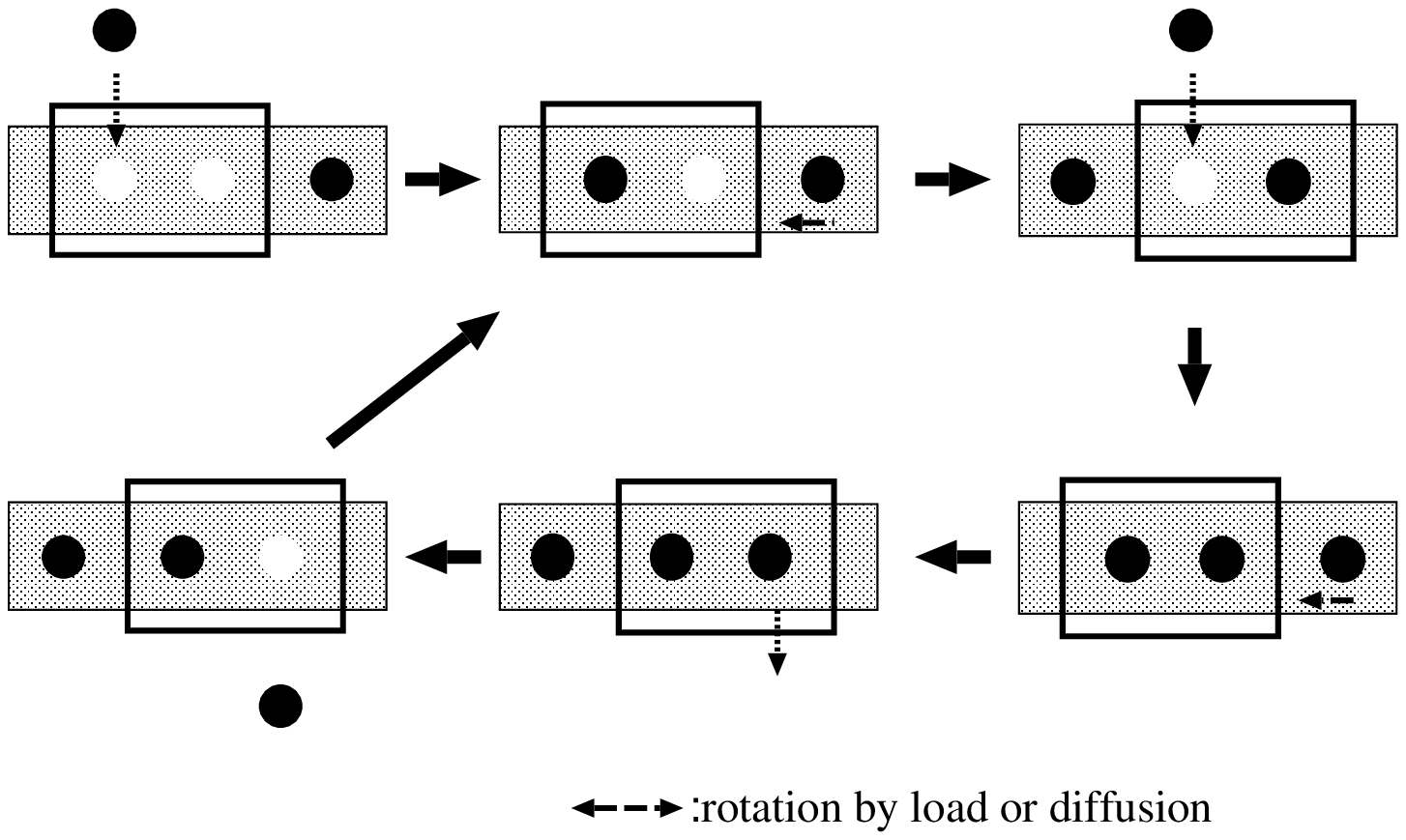}
\caption{A bad process. One proton pumping corresponds to more than
 one step of the rotation. The energy of the motion is wasted in this
 process.  \label{bad1}}
\includegraphics[width=7.5cm]{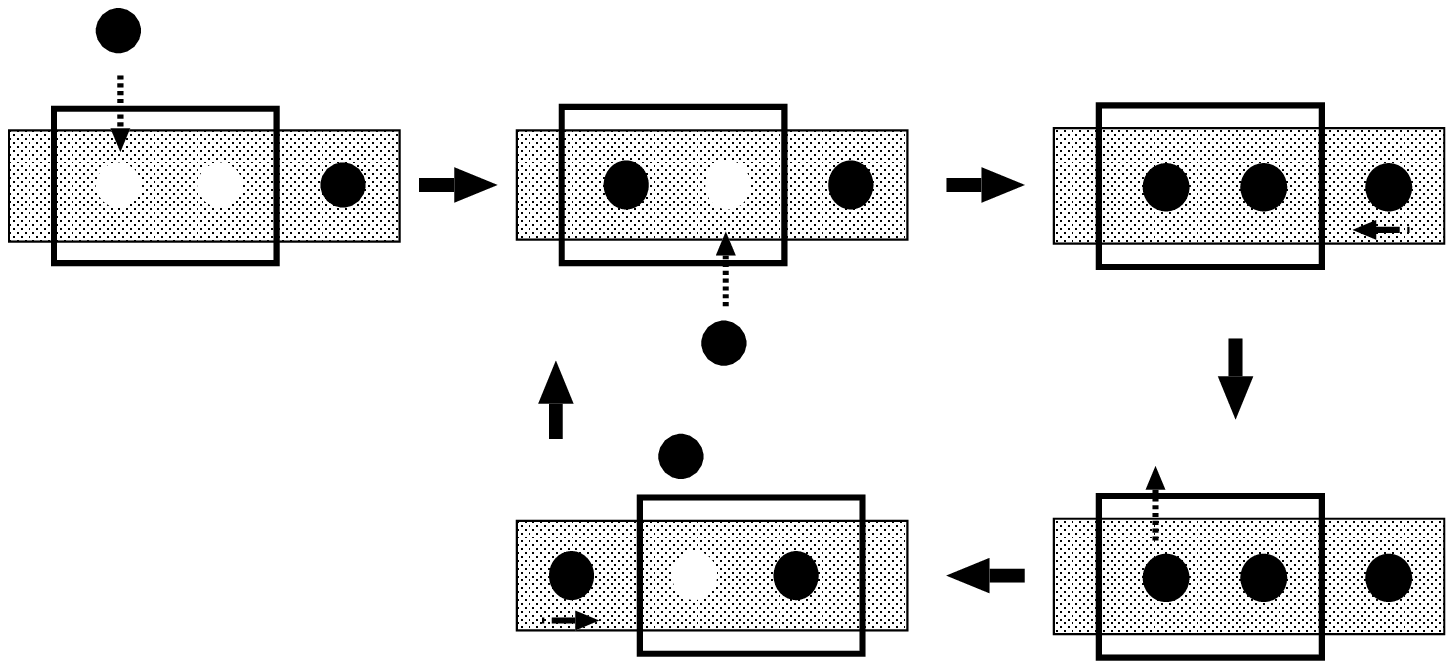}
\caption{Another bad process. A proton flows in the undesirable
 direction without rotation. This process is possible since the proton
 concentration of the outer side is kept higher than that of the inner
 side. This process suppresses the efficient energy transduction. 
 \label{bad2}}
\end{center}
\end{figure}

First of all, we give a qualitative consideration of the energy transduction, 
the coupling between the rotation and the proton flow. The good process
in the channel is shown in  FIG.\ref{good}: A proton binds to the left
binding site in the left figure. Then the {\bf c}-ring moves one step
leftward in the center figure. 
Finally the proton bound to the right binding site dissociates in the
right figure and 
the process returns back to the start. In this process one step of the
{\bf c}-ring rotation  corresponds to
one proton pumping. This is the efficient energy transduction. A bad
process is shown in FIG.\ref{bad1}. In the upper-right figure, a proton
binds to the 
left binding site before the proton bound to the right binding site
dissociates. Then in the lower-right figure, the {\bf c}-ring moves
leftward without proton dissociation. Therefore more than one step of
the rotation
corresponds to one proton pumping. In this process the energy is
transduced inefficiently. Another bad process is shown in FIG.\ref{bad2}. In
the upper-middle figure, a proton binds to the right binding site from
the acidic side before the rotation of the {\bf c}-ring. Next the {\bf c}-ring
rotates leftward in the upper-right figure. Finally the proton bound to
the left binding site dissociates then the {\bf c}-ring rotates rightward
due to diffusion. In
this process, a proton passes through the membrane along the gradient, in
the undesirable direction, without net step of the {\bf c}-ring. These bad
processes generate the loss of the energy transduction. For the higher efficiency, 
the good process should become more and the bad processes should become less.

\begin{figure}
\begin{center}
\includegraphics[width=7.5cm]{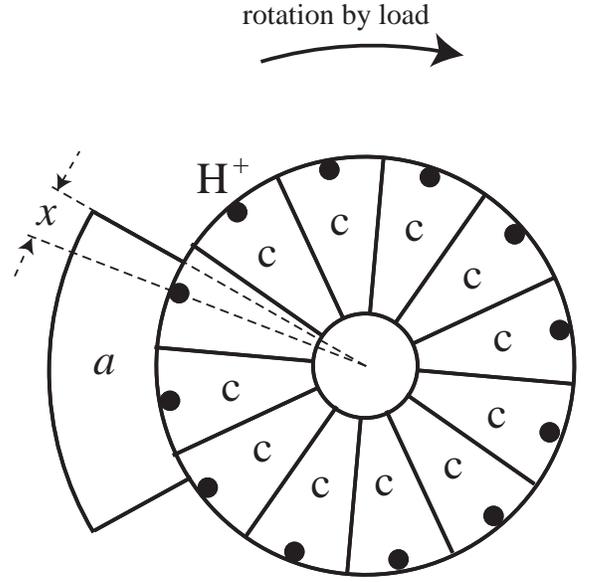}
\caption{The $F_o$ part viewed from the basic side. The position
 variable $x$ is defined as a rotation angle of the {\bf c}-ring. 
\label{coordinate}}
\end{center}
\end{figure}

The system is described by the Fokker-Planck equation, 
\begin{equation}
 \frac{\partial}{\partial t}{\bf{p}}(x,t)=-\frac{\partial}{\partial
 x}{\bf{\Pi}}(x,t)+{\hat{\bf K}}(x)\cdot{\bf{p}}(x,t) \label{fpeeq},
\end{equation}
\begin{equation}
 {\bf {\Pi}}(x,t)=\gamma{\hat{\bf f}}(x){\bf{p}}(x,t)-D\frac{\partial}
{\partial x}{\bf{p}}(x,t) \label{flow},
\end{equation}
\begin{eqnarray}
 {\bf{p}}(x,t)=\left[
\begin{array}{c}
p_{\rm E}(x,t)\\
p_{\rm R}(x,t)\\
p_{\rm L}(x,t) \\
p_{\rm F}(x,t)
\end{array}
\right],
\end{eqnarray}
\begin{eqnarray}
 {\bf{\Pi}}(x,t)=\left[
\begin{array}{c}
\Pi_{\rm E}(x,t)\\
\Pi_{\rm R}(x,t)\\
\Pi_{\rm L}(x,t) \\
\Pi_{\rm F}(x,t)
\end{array}
\right].
\end{eqnarray}      
Here ${\bf p}(x,t)$ is 4-component probability where $p_i(x,t)$ 
describes the probability that the {\bf c}-ring in state $i$ is at
position $x$ at time $t$. The state index $i$ refers to the state mentioned 
before; E, R, L, and F.  Similarly, ${\bf \Pi}$ describes the flow of
the probability. $\hat {\bf K}$ is the transition matrix which describes
changes between the states and will be given later.
$D$ and $\gamma$ are the diffusion constant and the friction constant,
respectively. They satisfy the Einstein's relation, $D=\gamma k_{\rm
B}T$ where $T$ is the temperature and $k_{\rm B}$ is the Boltzmann
constant. The matrix $\hat {\bf f}= {\rm diag}[\tau,\tau,\tau,\tau]$ is the
external load. The position variable  $x$ is defined as the rotation angle
with the origin placed at the center of the {\bf c}-ring and takes its
value from 0 to $\delta=2\pi/N$ where $N$ denotes the number of {\bf
c}-subunit consisting in the {\bf c}-ring. (See Fig.\ref{coordinate}.)
Hereafter, we fix $N=$12, which is the case of {\it E.Coli}. 
The position $x=0$ is defined as the situation where the left binding
site is at 
the left boundary of the channel and $x=\delta$ is as that where the right 
binding site is at the right boundary of the channel. 

As mentioned earlier, the conformational change of the
{\bf c}-subunit due to the protonation of the binding site is small.
Therefore we have assumed that all the state take the common values of the
diffusion constant $D$ and the friction constant $\gamma$. 
As an effect of the conformational change, we will only take
account of the hydrophobicity, which determines whether or not a {\bf
c}-subunit in the channel can move into the membrane.

The system has a periodicity in the sense that when one binding site in
the channel goes out and moves into the membrane, another binding site
comes into the channel. This periodicity and the hydrophobicity impose
the boundary conditions on the flows,
\begin{equation}
\Pi_{\rm E}(0,t)=\Pi_{\rm E}(\delta,t)=0,\label{nogoE}
\end{equation}
\begin{equation}
 \Pi_{\rm R}(0,t)=\Pi_{\rm L}(\delta,t)=0,\label{nogoRL}
\end{equation} 
\begin{equation}
\Pi_{\rm L}(0,t)=\Pi_{\rm R}(\delta,t), \label{peRL}
\end{equation}
and,
\begin{equation}
 \Pi_{\rm F}(0,t)=\Pi_{\rm F}(\delta,t). \label{peF}
\end{equation}
Similarly, the following boundary conditions of the probabilities are imposed, 
\begin{equation}
p_R(\delta)=p_L(0),\label{pRL}
\end{equation}
\begin{equation}
p_F(\delta)=p_F(0).\label{pF}
\end{equation}

The transition matrix ${\bf {\hat K}}$ is given as 
\[
{\bf {\hat K}}=
\]
\begin{equation}
\left[
\begin{array}{cccc}
-(k^{\rm R}_{\rm in}+k^{\rm L}_{\rm in})&k^{\rm R}_{\rm out}
&k^{\rm L}_{\rm out}&0\\
k^{\rm R}_{\rm in}&-(k^{\rm R}_{\rm out}+k^{\rm L}_{\rm in})
&0&k^{\rm L}_{\rm out}\\
k^{\rm L}_{\rm in}&0&-(k^{\rm R}_{\rm in}+k^{\rm L}_{\rm in})
&k^{\rm R}_{\rm out}\\
0&k^{\rm L}_{\rm in}&k^{\rm R}_{\rm in}
&-(k^{\rm R}_{\rm out}+k^{\rm L}_{\rm out})\\
\end{array}
\right],
\end{equation} 
where $k^i_j$($i$=R,L,$j$=in,out) denotes the rate at which a proton
binds to (in) or dissociate from (out) the right(R) or left(L) binding site.
Here we assumed that (1) the chemical reaction is sufficiently faster
than the motion of the {\bf c}-ring; (2) there is no proton hopping
between the binding sites and no correlation between the reactions; 
(3) the reaction rates are independent of the position, $x$. 

The steady state solution is of interest therefore the time variable is
omitted hereafter. Under these assumptions, this system
can be solved analytically. The matrix 
\begin{widetext}
 \begin{equation}
  \hat{\bf Q}=\left[
\begin{array}{cccc}
k^{\rm R}_{\rm out}k^{\rm L}_{\rm out}&-k^{\rm R}_{\rm out}&-k^{\rm L}_{\rm out}&1\\
k^{\rm R}_{\rm in}k^{\rm L}_{\rm out}&-k^{\rm R}_{\rm in}&k^{\rm L}_{\rm out}&-1\\
k^{\rm R}_{\rm out}k^{\rm L}_{\rm in}&k^{\rm R}_{\rm out}&-k^{\rm L}_{\rm in}&-1\\
k^{\rm R}_{\rm in}k^{\rm L}_{\rm in}&k^{\rm R}_{\rm in}&k^{\rm L}_{\rm in}&1\\
\end{array}
\right]
\end{equation}
diagonalizes the transition matrix $\hat {\bf K}$

 \begin{equation}
  \hat{\bf Q}^{-1}\hat{\bf K}\hat{\bf Q}=\left[
\begin{array}{cccc}
0&&&\\
&-(k^{\rm L}_{\rm in}+k^{\rm L}_{\rm out})&&\\
&&-(k^{\rm R}_{\rm in}+k^{\rm R}_{\rm out})&\\
&&&-(k^{\rm R}_{\rm in}+k^{\rm L}_{\rm in}+k^{\rm R}_{\rm out}+k^{\rm L}_{\rm out})\\
\end{array}
\right].
\end{equation}

Then Eq.(\ref{fpeeq}) in the steady state is reduced to
\begin{equation}
 0=-\frac{d}{dx}\left[\gamma\tau-D\frac{d}{dx}\right](\hat{\bf
 Q}^{-1}{\bf p}(x))
+\left[
\begin{array}{cccc}
0&&&\\
&-(k^{\rm L}_{\rm in}+k^{\rm L}_{\rm out})&&\\
&&-(k^{\rm R}_{\rm in}+k^{\rm R}_{\rm out})&\\
&&&-(k^{\rm R}_{\rm in}+k^{\rm L}_{\rm in}+k^{\rm R}_{\rm out}+k^{\rm L}_{\rm out})\\
\end{array}
\right](\hat{\bf Q}^{-1}{\bf p}(x)).\label{reduced}
\end{equation}
\end{widetext}
The solution is 
\begin{equation}
(\hat{\bf Q}^{-1}{\bf p}(x)) =\left[
\begin{array}{c}
C_1+C_2e^{\xi x}\\
C_3e^{\eta^+_{\rm L} x}+C_4e^{\eta^-_{\rm L} x}\\
C_5e^{\eta^+_{\rm R} x}+C_6e^{\eta^-_{\rm R} x}\\
C_7e^{\eta^+_{\rm LR} x}+C_8e^{\eta^-_{\rm LR} x}\\
\end{array}
\right],
\end{equation}
where $C_i$ ($i=1,2,\cdots,8$) are integral constants and
\begin{equation}
 \xi=\frac{\gamma\tau}{D}, \label{xi}
\end{equation}
\begin{equation}
 \eta_{\rm L}^{\pm}=\frac{\gamma\tau\pm\sqrt{(\gamma\tau)^2
+4D(k^{\rm L}_{\rm in}+k^{\rm L}_{\rm out})}}{2D}, \label{etal}
\end{equation}
\begin{equation}
 \eta_{\rm R}^{\pm}=\frac{\gamma\tau\pm\sqrt{(\gamma\tau)^2
+4D(k^{\rm R}_{\rm in}+k^{\rm R}_{\rm out})}}{2D}, \label{etar}
\end{equation}
\begin{equation}
 \eta_{\rm LR}^{\pm}=\frac{\gamma\tau\pm\sqrt{(\gamma\tau)^2
+4D(k^{\rm L}_{\rm in}+k^{\rm L}_{\rm out}+
k^{\rm R}_{\rm in}+k^{\rm R}_{\rm out})}}{2D} \label{etalr}.
\end{equation}
The integral constants $C_i$'s are determined by the boundary conditions
of the probabilities and their flows, Eqs.(\ref{nogoE})-(\ref{pF}).
To close the algebraic equations for $C_i$'s, the normalization condition
of the probability
\begin{equation}
\int^{\delta}_0 dx \sum_i p_i(x)=1,
\end{equation}
is needed.

Physical quantities which we will investigate are calculated as follows:
 First, the average rotation velocity $\langle v \rangle$ is given as,
\begin{equation}
\langle v \rangle =  \sum_i \bar{\Pi}_i,
\end{equation}
where
\begin{eqnarray}
\bar{\Pi}_i&=&\int^\delta_0 dx \Pi_i(x)
\nonumber
\\
&=&\gamma\tau \bar{p_i}+D[p_i(0)-p_i(\delta)],
\end{eqnarray}
and,
\begin{equation}
\bar{p_i}=\int^\delta_0 dx p_i(x).
\end{equation}
 Next, the proton pumping rate $N_{\rm p}({\rm H}^+)$ is written as, 
\begin{equation}
N_{\rm p}({\rm H}^+)=\frac{1}{2}(J_{\rm R}+J_{\rm L})
\end{equation}
where, 
\begin{equation}
J_{\rm R}=-k_{\rm out}^{\rm R}\bar{p_R}
    +k_{\rm in}^{\rm R}\bar{p_E}
    -k_{\rm out}^{\rm R}\bar{p_F}
    +k_{\rm in}^{\rm R}\bar{p_L},
\end{equation}
\begin{equation}
J_{\rm L} =k_{\rm out}^{\rm L}\bar{p_L}
    -k_{\rm in}^{\rm L}\bar{p_E}
    -k_{\rm in}^{\rm L}\bar{p_R}
    +k_{\rm out}^{\rm L}\bar{p_F}.
\end{equation}
Here $J_{\rm R}$ and $J_{\rm L}$ describe the proton flow of the right
and left binding site, respectively. 
Finally, let us define the efficiency of the energy
transduction\cite{Sekimoto}.  The
energy input per unit time is defined as the product of the external
load torque $\tau$ and the rotation velocity, $\langle v \rangle$. 
The energy output per unit time is the product of 
the free energy due to the transmembrane proton gradient, $\Delta G$,
 and the proton pumping rate, $N_{\rm p}({\rm H}^+)$. Therefore the
 efficiency  $e$  is defined  as, 
\begin{equation} 
e \equiv \frac{\Delta G \cdot N_{\rm p}({\rm H}^+)}
{\tau \cdot \langle v \rangle}.
\label{eff}
\end{equation}

\section{Results}
We investigate the dependence of the quantities, $\langle v \rangle$,
$N_{\rm p}({\rm H}^+)$, and $e$ on the transition rates, $\hat {\bf K}$. 
The transition rates are given by
\begin{equation}
\left[
\begin{array}{c}
k^{\rm R}_{\rm in}\\
k^{\rm L}_{\rm in}\\
k^{\rm R}_{\rm out}\\
k^{\rm L}_{\rm out}
\end{array} 
\right]
=10^K
\left[
\begin{array}{c}
10^{-pH_{\rm A}}e^{\phi/k_{\rm B}T}\\
10^{-pH_{\rm B}}e^{\phi/k_{\rm B}T}\\
10^{-pK_a}e^{-V/2k_{\rm B}T}\\
10^{-pK_a}e^{V/2k_{\rm B}T}\\
\end{array}
\right].\label{chem}
\end{equation}
The parameters used for the calculation are, according to Ref.\cite{EWO},
 given in TABLE.\ref{pars}. They are plausible values in living things.
\begin{table}
\begin{center}
\caption{Parameters used for calculation.\label{pars}}
\begin{tabular}{|p{10em}|c|p{13em}|}
\hline
diffusion constant of the {\bf c}-ring&D&$2 \times 10^4$sec${}^{-1}$\\
drag constant of the {\bf c}-ring&$\gamma$&$5\times10^3$sec${}^{-1}
$pN${}^{-1}$nm${}^{-1}$\\
temperature&$k_{\rm B}T$&$4$pN$\cdot$nm\\
proton concentration in acidic side&$pH_A$&$6.6$\\
proton concentration in basic side&$pH_B$&$7.6$\\
surface effect&$\phi$&$2.0k_{\rm B}T$\\
membrane potential&$V$&$5.6k_{\rm B}T$\\
\hline
acidity of binding site&$pK_a$&control parameter\\
external load&$\tau$&control parameter\\
\hline
\end{tabular}
\end{center}
\end{table}
In our previous work \cite{MSK}, we pointed out that the inequalities
 between the chemical reaction rates, $k^i_j$'s, each other and the
 reciprocal of the relaxation time of the system, $T_{\rm relax}^{-1}=
(\gamma\tau)^2/D$, affect the behavior of the physical quantities in the
 motor process. 
 It is natural to expect that these inequalities also play important
 roles in the pump process. The chemical reaction rates depend on many
 kinds of factor, {\it e.g.} the diffusion constant of the proton, the
 acidity of the proton binding site, the concentrations of proton of
 both sides, {\it etc}. The overall factor $10^K$ means  the effective
 proton absorption rate of the path, which is
 regarded as common for all the transition rates. 

The free energy accompanying the proton translocation due to the
transmembrane proton gradient is given as
\begin{equation}
\Delta G=V+k_{\rm B}T \ln [10^{\Delta pH}]
\nonumber
\end{equation}
\begin{equation}
=31.6[{\rm pN} \cdot {\rm nm}], 
\end{equation}
where $\Delta pH=pH_B-pH_A$. The detailed balance is violated by this
free energy and the pumping is against this energy.  

We investigate in detail two sets of inequalities between the transition
rates;  case 1)
$k_{\rm out}^{\rm L} > k_{\rm out}^{\rm R} > k_{\rm in}^{\rm R} >
 k_{\rm in}^{\rm L}$
 and case 2)  
$k_{\rm out}^{\rm L} > k_{\rm in}^{\rm R} > k_{\rm out}^{\rm R} >
 k_{\rm in}^{\rm L}$.
 They are derived from the parameters in TABLE.\ref{pars}, 
 and the condition under which our model works well as a motor. 
 We take the
 value of the acidity of proton binding site, $pK_a$=4.5 and 5.5 as
 typical value which gives the set of inequalities 1) and 2),
 respectively. 

The external load $\tau$ is also regarded as a
 controllable parameter, since the load has not been so far measured
 experimentally in the pump process we discuss here. 

\begin{figure}
\begin{center}
\includegraphics[width=7.5cm]{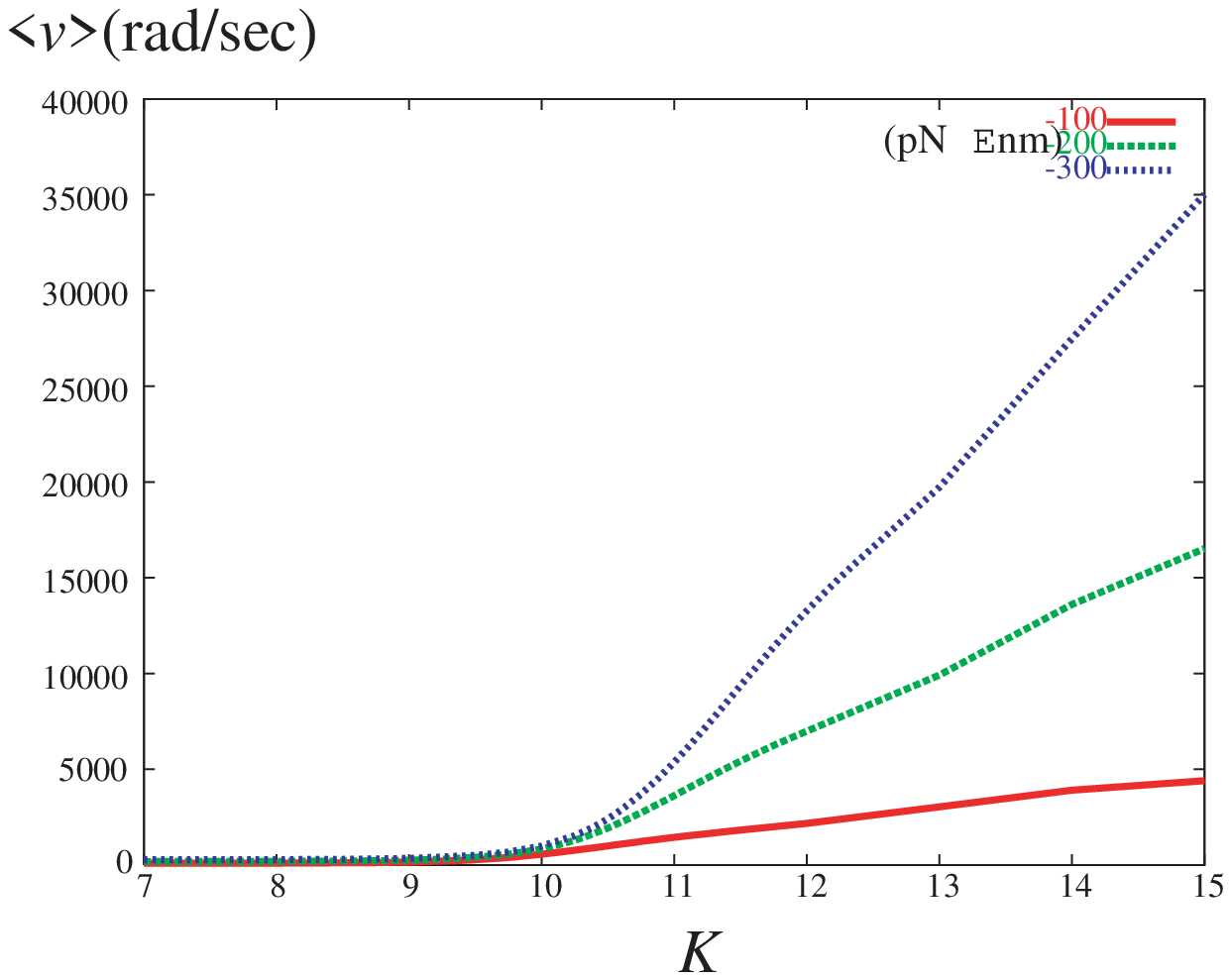}
\caption{The rotation velocity $\langle v \rangle$ in the case of the
 set of inequalities 
$k_{\rm out}^{\rm L} > k_{\rm out}^{\rm R} > k_{\rm in}^{\rm R} >
 k_{\rm in}^{\rm L}$ ($pK_a$=4.5, case 1)).
\label{v45}}
\includegraphics[width=7.5cm]{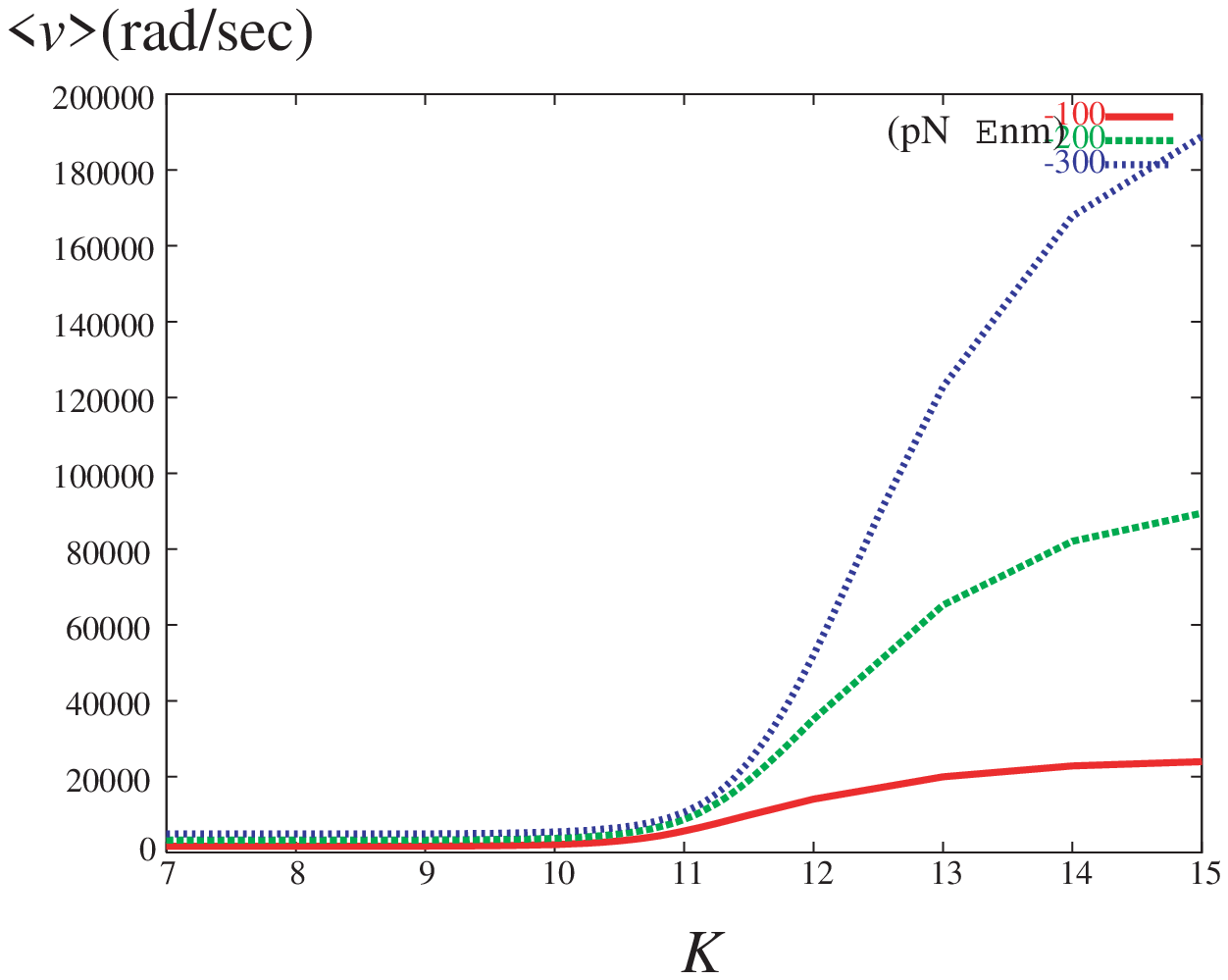}
\caption{The rotation velocity $\langle v \rangle$ in the case of the
 set of inequalities  
$k_{\rm out}^{\rm L} > k_{\rm in}^{\rm R} > k_{\rm out}^{\rm R} >
 k_{\rm in}^{\rm L}$ ($pK_a$=5.5, case 2)).
 \label{v55}}
\end{center}
\end{figure}

\begin{figure}
\begin{center}
\includegraphics[width=7.5cm]{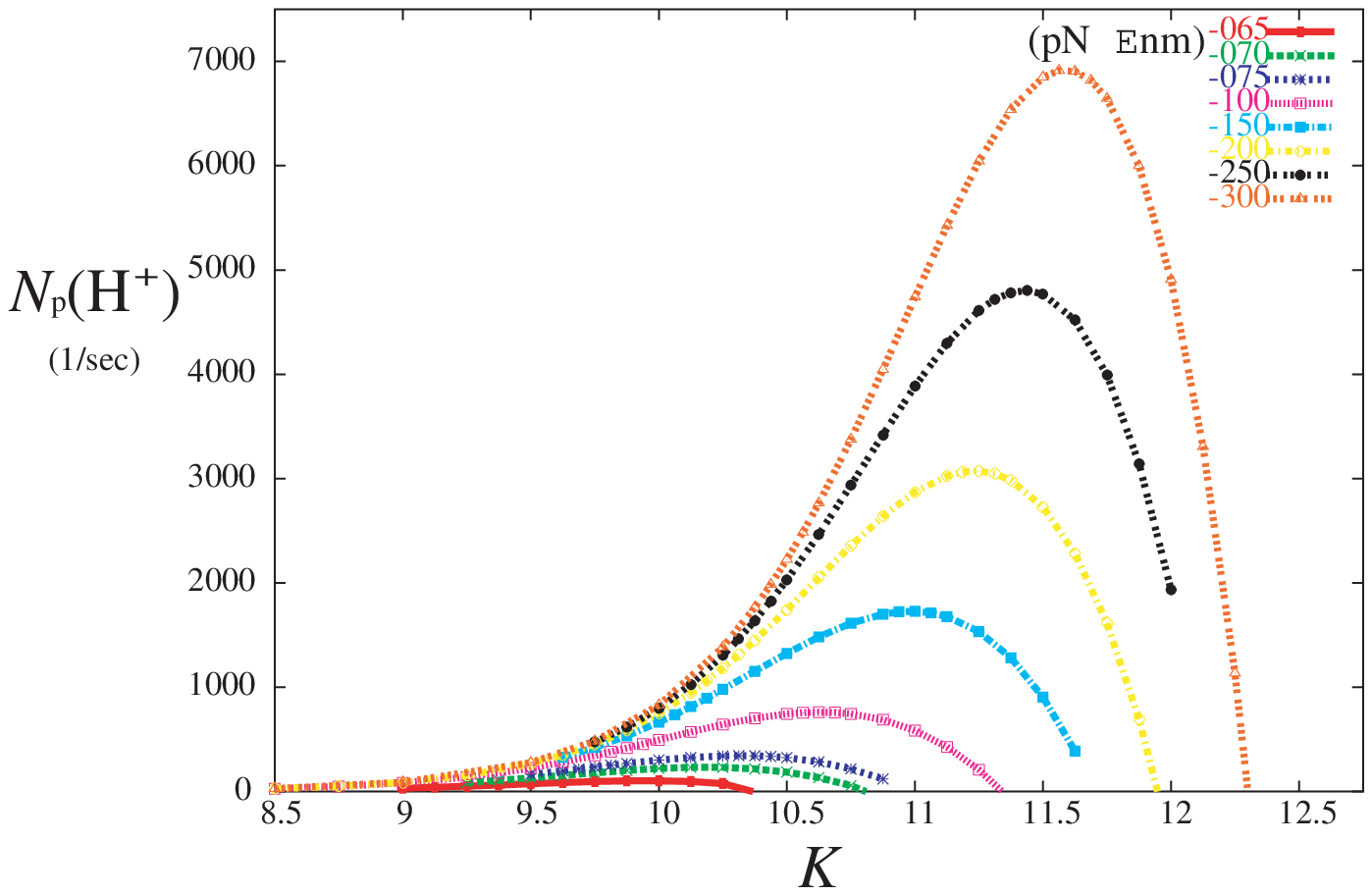}
\caption{The proton pumping rate $N_{\rm p}({\rm H}^+)$ in the case 1).
 For fixed load $\tau$, this quantity 
 has a peak at about $k_{\rm out}^{\rm L} \sim T_{\rm relax}^{-1} \equiv 
(\gamma \tau)^2/D$. \label{nh45}}
\includegraphics[width=7.5cm]{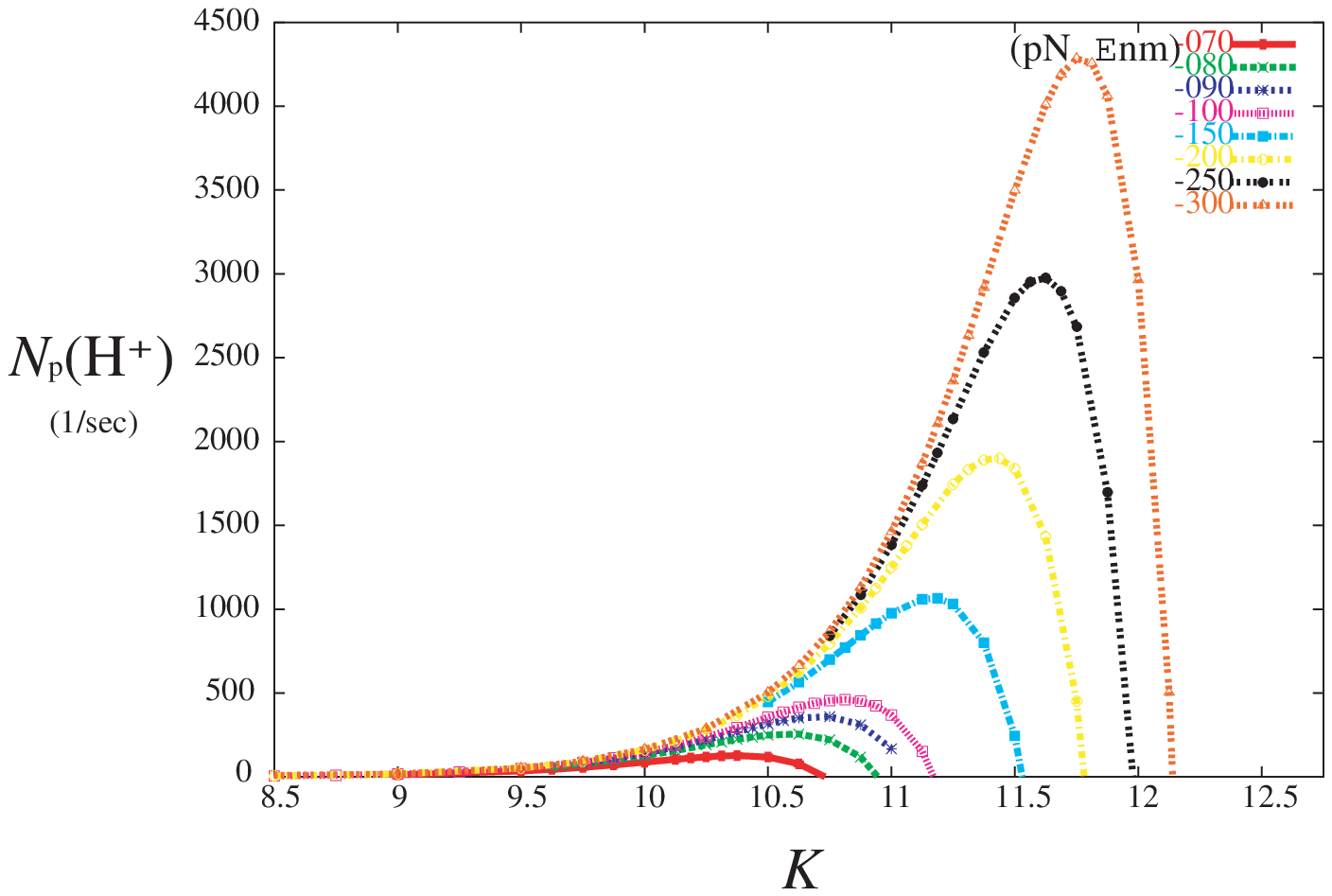}
\caption{$N_{\rm p}({\rm H}^+)$ in the case 2). \label{nh55}}
\end{center}
\end{figure}

\begin{figure}
\begin{center}
\includegraphics[width=7.5cm]{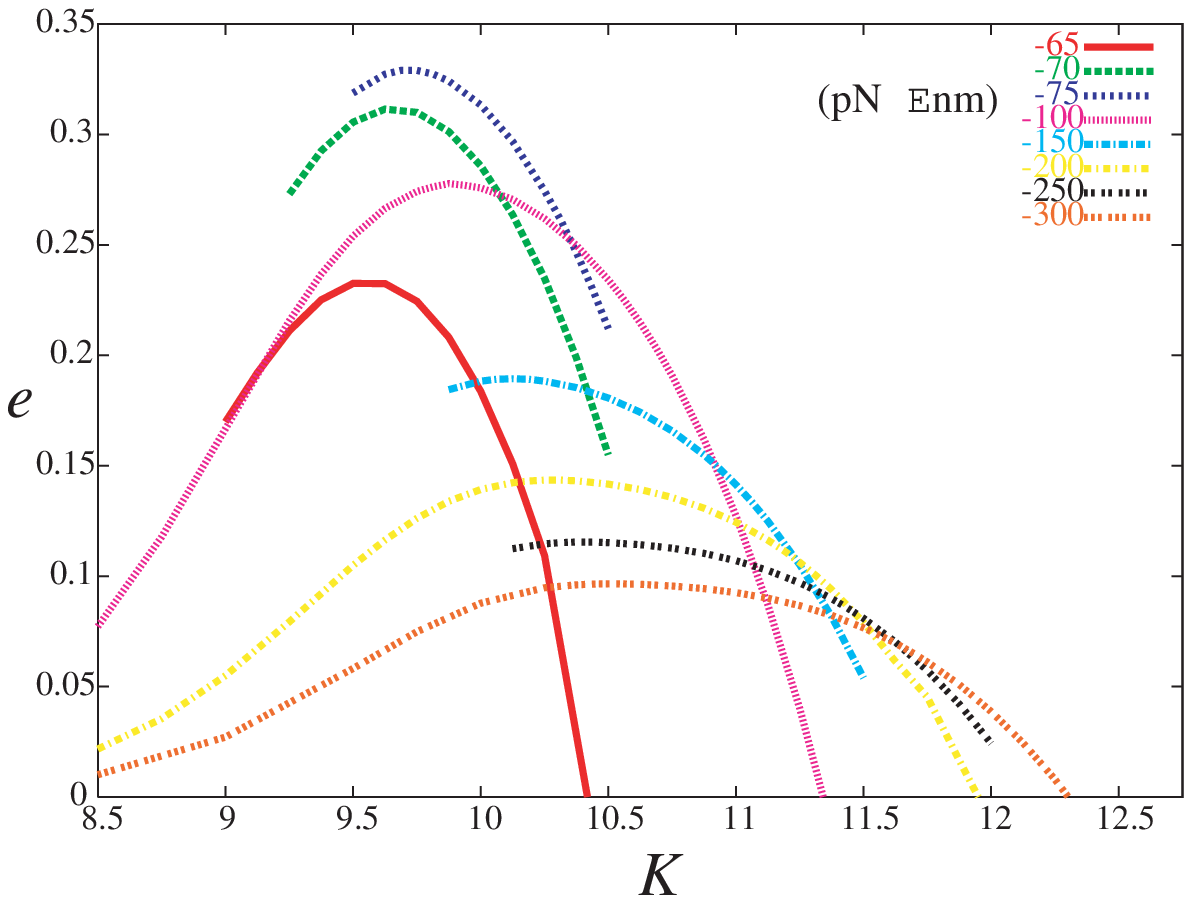}
\caption{The energy transduction efficiency $e$ in the case 1).
For a fixed load $\tau$, there is a
 peak at about $T_{\rm relax}^{-1} \gtrsim k_{\rm out}^{\rm L} >
 T_{\rm load}^{-1}$. \label{e45}}
\includegraphics[width=7.5cm]{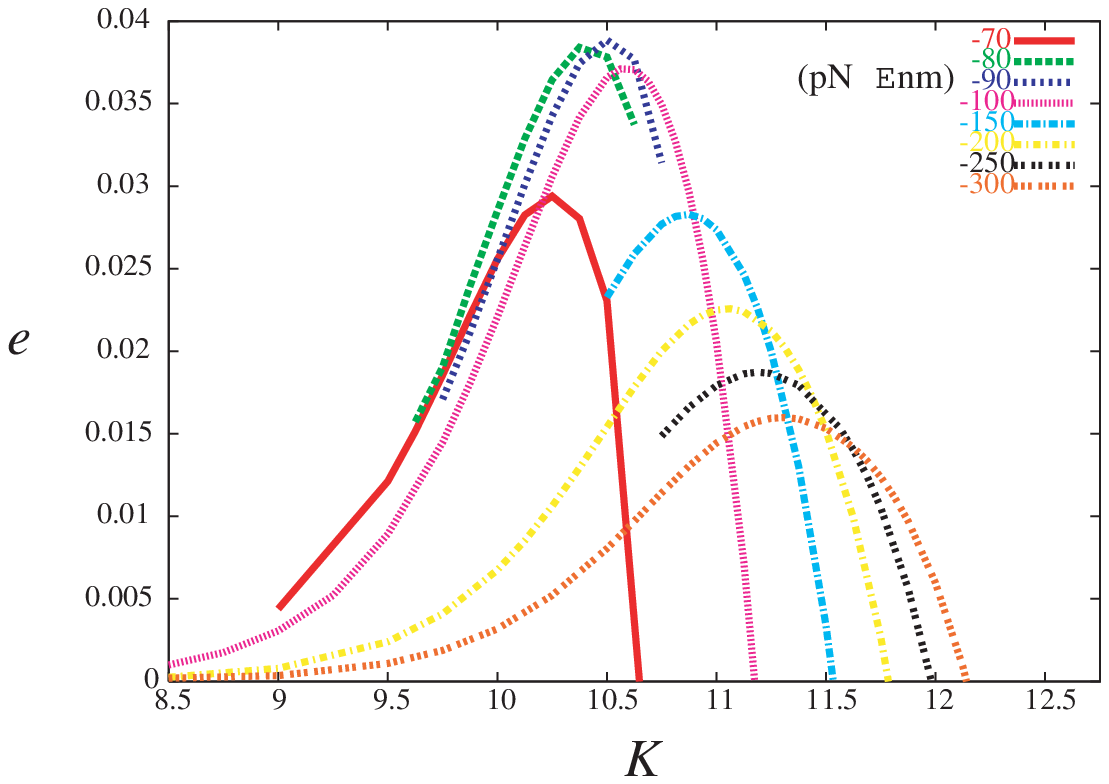}
\caption{The energy transduction efficiency $e$ in the case 2). For a
 fixed load $\tau$, there is a 
peak at about $T_{\rm relax}^{-1} \gtrsim k_{\rm out}^{\rm L} > 
T_{\rm load}^{-1}$. In this case the value is about one
 order smaller than the former case. \label{e55}}
\end{center}
\end{figure} 
The rotation velocity $\langle v \rangle $ is plotted in FIGs.\ref{v45}
 and \ref{v55} in the case 1) and 2), respectively. There is no
 qualitative difference between two cases. There is almost no rotation
 when $K$ is small, {\it i.e.} the chemical reaction rates are low.
 Then it monotonically increases and finally saturates at $K>$15, which
 is out of the figures. As expected, it becomes faster for the larger load.

The proton pumping rate $N_{\rm p}({\rm H}^+)$ is plotted in
 FIGs.\ref{nh45} and 
 \ref{nh55}. Almost no pumping is observed for small $K$ in both cases.
 Then it increases as $K$ increases. It reaches a peak at a certain
 finite value of $K$, such that $k_{\rm out}^{\rm L} \sim T^{-1}_{\rm
 relax}$, and then falls down abruptly. The larger the load, the more
 the peak value increases. 
        
\begin{figure}
\begin{center}
\includegraphics[width=7.5 cm]{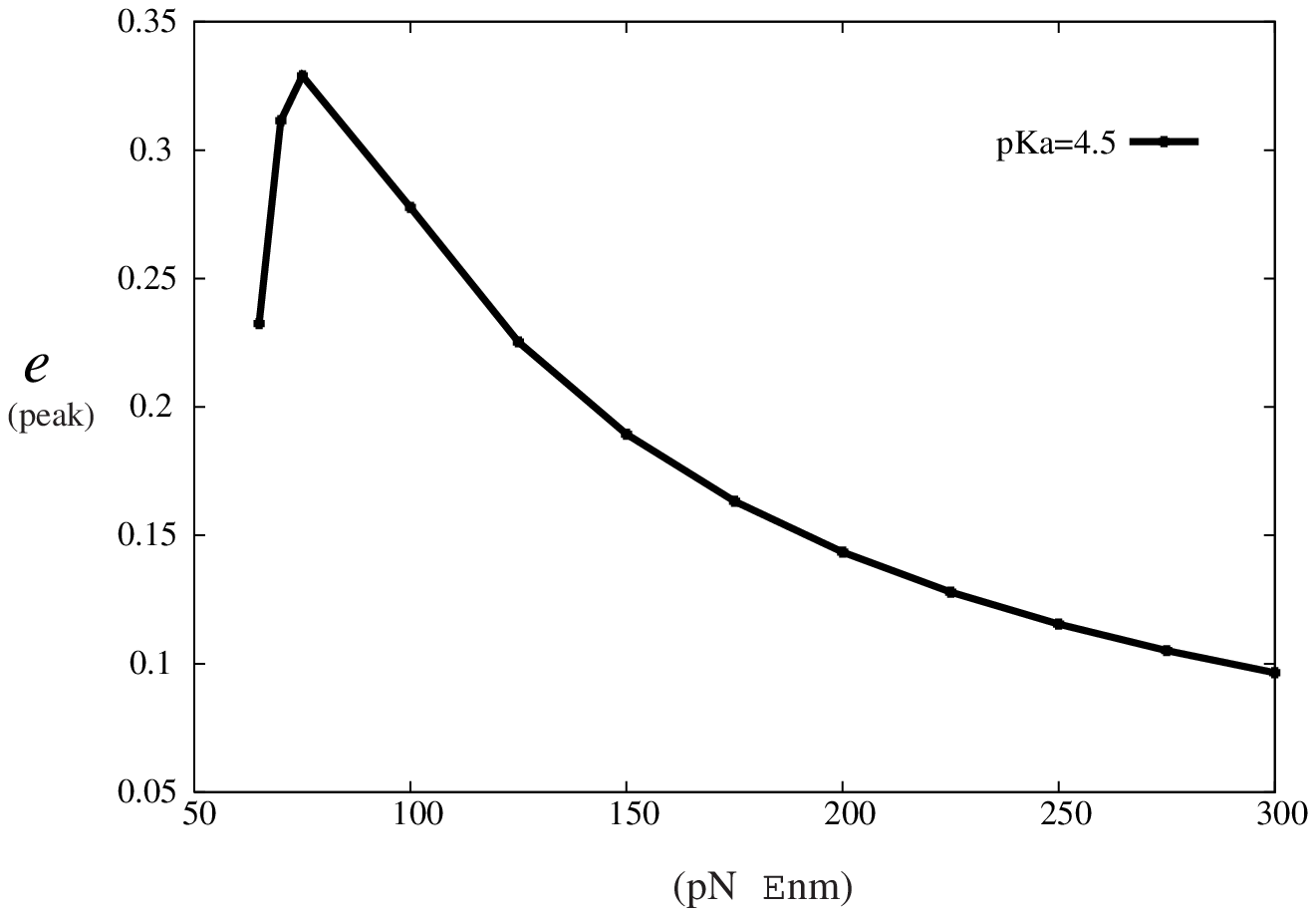}
\caption{The variation of the peak value of the  energy transduction
 efficiency  $e$ in the case 1) with respect to the load $\tau$. 
Below the critical load $\tau_c \sim $60(pN$\cdot$nm), the system
 cannot work as a pump. 
It has a peak for a certain finite value of $\tau$, 
$\tau_{\rm opt} \sim$75(pN$\cdot$nm). 
 \label{epeak45}}
\includegraphics[width=7.5cm]{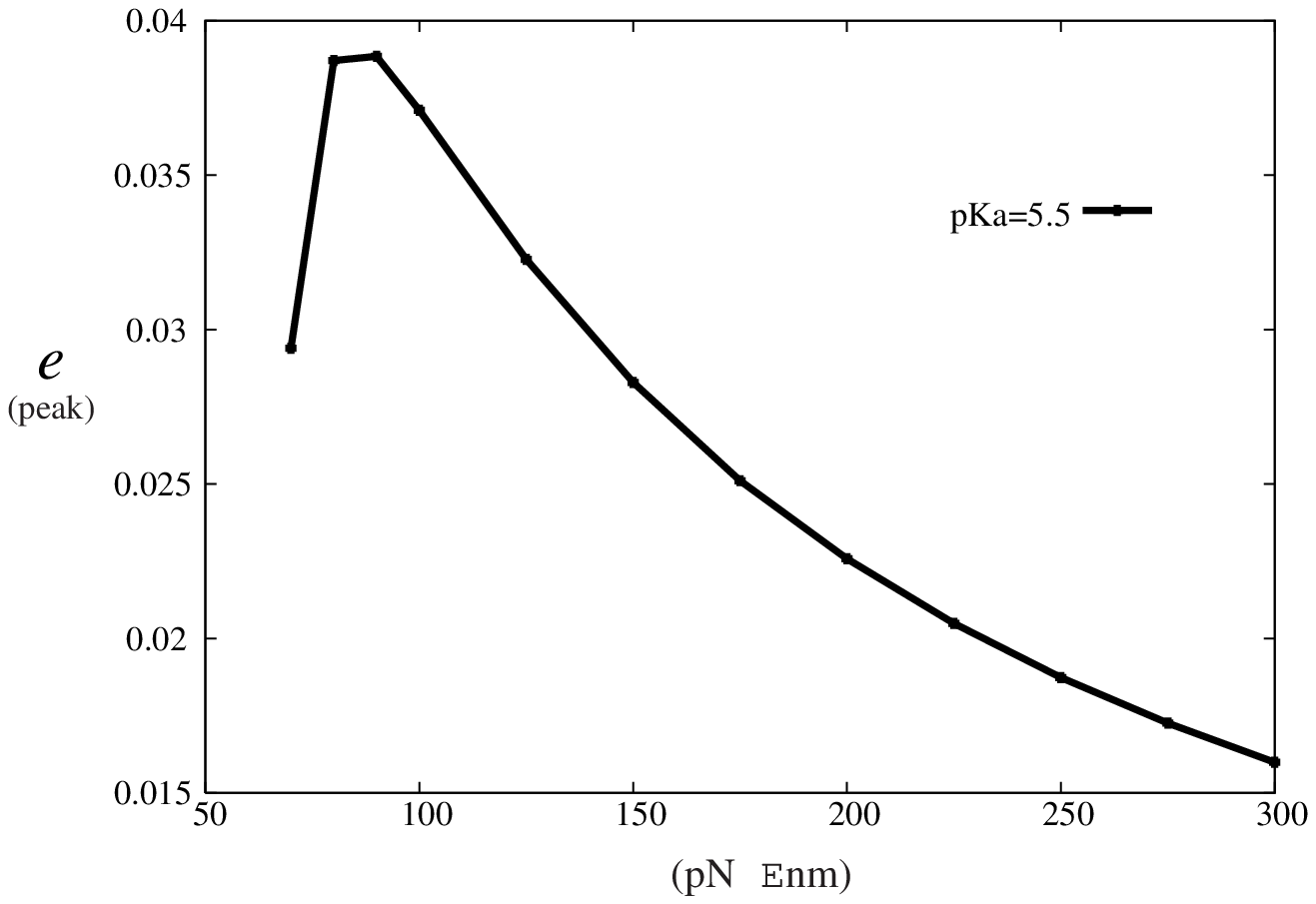}
\caption{The variation of the peak value of the energy transduction
 efficiency  $e$ in the case 2) with respect to the load $\tau$. 
For $\tau < \tau_c \sim$70(pN$\cdot$nm), the system cannot work as a pump.
There is a peak for a certain finite value of 
 $\tau$, $\tau_{\rm opt} \sim$90(pN$\cdot$nm). The value is
 about one order smaller than the case 1). \label{epeak55}}
\end{center}
\end{figure}

The energy transduction efficiency $e$ is plotted in FIGs.\ref{e45} and
 \ref{e55}. For respective fixed loads $\tau$,
 it has a peak for a certain finite
 value of $K$. This fact reflects the above behaviors of the rotation
 velocity  and the proton pumping rate. The peak is placed on $K$ such
 that $T_{\rm relax}^{-1} \gtrsim k_{\rm out}^{\rm L} > T_{\rm
 load}^{-1}$ holds. Here $T_{\rm load} \equiv \delta / \gamma \tau$
 is the characteristic time that it takes for the {\bf c}-ring to rotate
 one step leftward by the load. 
 The variation of the peak value with respect to
 the load $\tau$ is shown in FIGs.\ref{epeak45} and \ref{epeak55}. 
 For the variation of the load
 $\tau$, there exists an optimal value of $\tau$ such that the peak
 value of the efficiency becomes maximum. In
 the case 1), $\tau \sim$75pN$\cdot$nm and $\tau\sim$90pN$\cdot$nm in
 the case 2). Note that the efficiency in the case 1) is about one order
 larger than that in the case 2). Therefore, the pump mechanism works
 most efficiently when the set of inequalities    
$ T_{\rm relax}^{-1} \gtrsim k_{\rm out}^{\rm L} >
T_{\rm load}^{-1} > k_{\rm out}^{\rm R} > k_{\rm in}^{\rm R} >
 k_{\rm in}^{\rm L}$
holds and the optimal value of $\tau \sim$75pN$\cdot$nm is fulfilled.

We can understand qualitatively the reason why the above inequalities
 are required for the efficient pump process. Firstly, the inequality
 $T_{\rm relax}^{-1} \gtrsim k^{\rm L}_{\rm out}$ is required to suppress the bad
 process shown in FIG.\ref{bad2}. If the inequality is not satisfied,  
 the left binding site in the channel becomes unprotonated before the
 relaxation by the fluctuation hence it cannot move into the
 membrane. As a result the bad process in FIG.\ref{bad2} can occur
 easily. Secondly, the
 inequality $k^{\rm L}_{\rm out} > T_{\rm load}^{-1}$ indicates that 
the bad process shown in FIG.\ref{bad1}, rotation without proton
pumping, rarely occurs. Thirdly, the inequality $k_{\rm out}^{\rm R} > k_{\rm
in}^{\rm R}$, which distinguishes the case 1) from 2), is needed to prevent
the {\bf c}-ring from moving rightward.

For $\tau_{\rm opt}$, we note that it is determined in highly nontrivial
manners. To see this, let us go back to Eq.(\ref{eff}). As we see in
FIGs.\ref{v45}-\ref{nh55}, both the numerator of Eq.(\ref{eff}), which
is proportional to the proton pumping rate $N_{\rm p}({\rm H}^+)$, and
the denominator, which is proportional to the rotation velocity $\langle
v \rangle$ and the load $\tau$, increase monotonically as the load
increases. Therefore, the subtle difference between the increasing
rates of them determines $\tau_{\rm opt}$.   

\section{summary and discussion}
We have investigated the pump process of the ratchet model of the
rotatory molecular motor. The model can be solved analytically and the
physical quantities such as the rotation velocity and the proton pumping
rate are calculated. The energy transduction efficiency is obtained from
these physical quantities. 
The best efficiency condition is that the set of inequalities 
$ T_{\rm relax}^{-1} \gtrsim k_{\rm out}^{\rm L} > T_{\rm load}^{-1} 
> k_{\rm out}^{\rm R} > k_{\rm in}^{\rm R} > k_{\rm in}^{\rm L}$ 
holds under an optimal value of the load $\tau_{\rm opt}$. 

We compare the result obtained here with that for the motor process. In
Ref.\cite{MSK} we find that the most efficient condition for the motor
process is that the set of inequalities 
$k_{\rm out}^{\rm L} >  T_{\rm relax}^{-1} > k_{\rm out}^{\rm R} 
> k_{\rm in}^{\rm R} > k_{\rm in}^{\rm L}$ holds, under the experimentally
reported load, $\tau \sim$ 40pN$\cdot$nm \cite{NYY}. 
The set of inequalities required between the chemical reaction rates
coincide with that for the pump process obtained here. Therefore the
$F_o$ part can work both as a motor and a pump efficiently by tuning the
load $\tau$. In this sense these conditions
are consistent.  

As far as we know, experiments which directly measures the physical
quantities, such as the rotation velocity and the pumping rate,
have not been reported so far. However an experiment has been reported that
in a whole $F_oF_1$ ATP synthase the conformational change of one subunit
of $F_1$ part suppresses electronically and/or sterically the rotation
hydrolysing ATP\cite{spt}. This effect may be interpreted to vary the load
effectively and control which process occurs.    

In the model we considered here, only the diffusion process is taken
into account. The highest efficiency obtained is $e \sim$0.3, and it seems
to be not high enough, since molecular
motors are thought to transduce energy almost without loss
\cite{NYY}. There are several possibilities to solve this problem. The
first is to reconsider the definition of the efficiency
(\ref{eff}), since our definition, following the argument in
Ref.\cite{Sekimoto}, is different from that in Ref.\cite{NYY}. The
second is to reconsider the relations between physical quantities
appropriate near the equilibrium, such as the Einstein's relation, since
the system is far from equilibrium. The third is to introduce another
mechanism, such as the electrostatic interaction between residues in
Ref.\cite{EWO}. These may lead to our better understanding of the
energetics of non-equilibrium systems.

\end{document}